\documentclass{ifacconf}
\usepackage{filecontents}

\makeatletter
\let\old@ssect\@ssect % Store how ifacconf defines \@ssect
\makeatother
\usepackage{natbib}  
\usepackage{hyperref}
\makeatletter
\def\@ssect#1#2#3#4#5#6{%
	\NR@gettitle{#6}% Insert key \nameref title grab
	\old@ssect{#1}{#2}{#3}{#4}{#5}{#6}% Restore ifacconf's \@ssect
}
\usepackage{graphicx}      % include this line if your document contains figures
\usepackage{amsmath,amssymb,amsfonts}
\usepackage{algorithm}
\usepackage{algpseudocode}
\usepackage{graphicx}
\usepackage{textcomp}
\usepackage{xcolor}
\usepackage{stfloats}
\usepackage{booktabs}
\usepackage{enumitem}
\usepackage{fixltx2e}

\begin{document}
\begin{frontmatter}
% \title{Parallel Subspace Identification and ARX Modeling Using Weighted Least-Squares} 
\title{Weighted Least-Squares PARSIM} 
\thanks[footnoteinfo]{This work was supported by VINNOVA Competence Center AdBIOPRO, contract [2016-05181] and by the Swedish Research Council through the research environment NewLEADS (New Directions in Learning Dynamical Systems), contract [2016-06079], and contract 2019-04956.}

\author{Jiabao He, Cristian R. Rojas and H\r{a}kan Hjalmarsson} 
 
\address{Division of Decision and Control Systems, School of Electrical Engineering and Computer Science, KTH Royal Institute of Technology, 100 44 Stockholm, Sweden(e-mail: jiabaoh, crro, hjalmars@kth.se).}

\begin{abstract}                % Abstract of 50--100 words
Subspace identification methods (SIMs) have proven very powerful for estimating linear state-space models. To overcome the deficiencies of classical SIMs, a significant number of algorithms has appeared over the last two decades, where most of them involve a common intermediate step, that is to estimate the range space of the extended observability matrix. In this contribution, an optimized version of the parallel and parsimonious SIM (PARSIM), PARSIM\textsubscript{opt}, is proposed by using weighted least-squares. It not only inherits all the benefits of PARSIM but also attains the best linear unbiased estimator for the above intermediate step. Furthermore, inspired by SIMs based on the predictor form, consistent estimates of the optimal weighting matrix for weighted least-squares are derived. Essential similarities, differences and simulated comparisons of some key SIMs related to our method are also presented. 
\end{abstract}


\end{frontmatter}
%===============================================================================

\section{Introduction}
Subspace identification methods (SIMs) are attractive for their numerical robustness and general parametrization for multiple-input multiple-output (MIMO) systems. As summarized in \cite{Qin2006overview}, most classical SIMs can be unified into the theorem proposed in \cite{Van1995unifying}, such as canonical variate analysis (CVA) \citep{Larimore1990canonical}, N4SID \citep{Van1994n4sid}, subspace splitting \citep{Jansson1996linear} and MOESP \citep{Verahegen1992subspace}. To be specific, they involve the following steps: First, a high-order model is estimated by regression or projection. Second, the previous high-order model is reduced to an observable low-dimensional subspace using weighted singular value decomposition (SVD). Third, a balanced realization of state-space matrices is obtained based on the structure of the reduced observability matrix and the framework of maximum likelihood (ML). Despite the tremendous development of SIMs in both theory and practice, some drawbacks of classical SIMs should be emphasized \citep{Qin2006overview}. The first one is that the model format used in SIMs during the above projection step is non-causal. As a result, the estimated parameters have inflated variance due to the fact that extra and unnecessary terms are included in the model. The second one is that some SIMs are biased for closed-loop data, which require special treatments. Third, due to that SIMs are composed of several steps, including the projection and the weighted SVD step, it is difficult to analyze their statistical properties. There are some significant contributions on this problem, such as \cite{Bauer2005asymptotic,Bauer2000analysis,Chiuso2004asymptotic,Chiuso2005consistency}, but a complete statistical analysis and comparison of different weighting matrices regarding optimality is still unavailable. For example, the question of whether there are subspace methods that are asymptotically efficient is still unresolved some 50 years after this family of methods was introduced. Furthermore, SIMs are generally not believed to be as accurate as the prediction error method (PEM).

To address those problems, especially the first two, a significant number of SIMs has appeared over the last two decades. For a comprehensive overview of these methods, we refer to \cite{Qin2006overview}, \cite{Chiuso2007relation} and \cite{van2013closed}. This paper concentrates on the first problem. In particular, we propose an optimized version of the parallel and parsimonious SIM (PARSIM) \citep{Qin2005novel}. In addition to inheriting all the benefits of PARSIM, the problem of estimating the range space of the extended observability matrix is fitted into the ML framework. As a result, it attains the best linear unbiased estimator (BLUE) and subsequently implies smaller variances for estimating system matrices.

Besides standing on the shoulders of PARSIM, our method is also inspired by other closed-loop SIMs, such as subspace and ARX modeling (SSARX) \citep{Jansson2003subspace}, and SIMs based on the predictor identification (PBSID) \citep{Chiuso2005consistency, Chiuso2007role} and its optimized version PBSID\textsubscript{opt} \citep{Chiuso2007relation}. In the following section, we will integrate models and formulations to illustrate essential similarities and differences among those methods, along with how they relate to our method. In Section \ref{Sct3}, the implementation of our method is presented. Some numerical examples are provided in Section \ref{Sct4} to compare our method with the state-of-the-art, which suggest that our method should be considered as one of the most appealing SIMs. The paper is then concluded in Section \ref{Sct5}.

\section{Preliminaries} \label{Sct2}
\vspace{-3mm}
\subsection{Problem Formulation and Assumptions} \label{Sct2.1}

Consider the following discrete linear time-invariant (LTI) system on innovations form:
\begin{subequations} \label{E1}
	\begin{align}
		x_{k + 1} &= Ax_{k}  + Bu_{k} + Ke_{k}, \label{E1a}\\
		y_{k} &= Cx_{k} + Du_{k} +e_{k}, \label{E1b}		
	\end{align}
\end{subequations}
where $x_{t}\in \mathbb{R}^{n_x}$, $u_{t}\in \mathbb{R}^{n_u}$, $y_{t}\in \mathbb{R}^{n_y}$ and $e_{t}\in \mathbb{R}^{n_y}$ are the state, input, output and innovations, respectively. We are interested in estimating the system matrices $A$, $B$, $C$, $D$ and the Kalman gain $K$ using input and output data. System \eqref{E1} can be represented in predictor form as
\begin{subequations} \label{E2}
	\begin{align}
		x_{k + 1} &= \bar Ax_{k}  + \bar Bu_{k} + Ky_{k}, \label{E2a}\\
		y_{k} &= Cx_{k} + Du_{k} + e_{k}, \label{E2b}		
	\end{align}
\end{subequations}
where $\bar A = A-KC$ and $\bar B= B-KD$.

As pointed out in \cite{Qin2006overview}, most SIMs use one of the above forms. The advantage of the predictor form over the innovations form is that $\bar A$ is guaranteed to be stable even if $A$ is unstable, which facilitates the estimation of both stable and unstable processes. However, the drawback is that, for a finite number of samples, the optimal Kalman gain $K$ is time-varying, which results in a time-varying $\bar A$ even though $A$ is time-invariant.

Next, we recap the state-of-the-art in SIMs to expose their essential similarities and differences. Since their key variations are rooted in obtaining Markov parameters and the range space of the extended observability matrix, while the remaining steps to estimate the model parameters are similar, we will only cover these steps. Before proceeding further, we introduce the following assumptions  commonly used in SIMs:

\begin{assum} \label{Asp1}
	\begin{enumerate} %[label=(\alph*)]
		\item The system is stable, i.e., the eigenvalues of $A$ are strictly inside the unit disk.
		\item The system is minimal, i.e., $(A,[B,K])$ is controllable and $(A,C)$ is observable.
		\item The innovations sequence $\{e_k\}$ is a stationary, zero-mean, white noise with $\mathbb{E}(e_ke_l) = R\delta_{kl}$, where $\delta_{kl}$ is the Kronecker delta. For simplicity,  we assume $R = \sigma_e^2I$ throughout the paper, but we believe that our method can be adapted to arbitrary covariance matrices $R$.
		\item The input sequence $\{u_k\}$ is quasi-stationary and independent of $\{e_k\}$.
		\item The input $\{u_k\}$ is persistently exciting of order $f + p$, where the future and past horizons $f$ and $p$ will be defined later.
	\end{enumerate}
\end{assum}

\subsection{SIMs Based on Innovations Form} \label{Sct2.2}
\vspace{-3mm}
Here we provide a short review on classical SIMs using the innovations form \eqref{E1}, with focus on PARSIM. An extended state-space model for \eqref{E1} can be derived as
\begin{subequations} \label{E3}
	\begin{align}
		Y_f &= \Gamma_fX_k + G_fU_f + H_fE_f, \label{E3a}\\
		Y_p &= \Gamma_pX_{k-p} + G_pU_p + H_pE_p, \label{E1b}		
	\end{align}
\end{subequations}
where $f$ and $p$ denote future and past horizons, respectively. Also, the extended observability matrix is
\begin{equation} \label{E4}
	\Gamma_f = \begin{bmatrix}
		{{C^{\top}}}&{{{\left( {CA} \right)}^{\top}}}& \cdots &{{{\left({C{A^{f - 1}}} \right)}^{\top}}}
	\end{bmatrix}^{\top},
\end{equation}
and $G_f$ with $H_f$ are Toeplitz matrices of Markov parameters with respect to the input and innovations, 
\begin{subequations} \label{E5}
	\begin{align}
		G_{f} &= \begin{bmatrix}
			{D}&{0}& \cdots &0\\
			{CB}&D& \cdots &0\\
			\vdots & \vdots & \ddots & \vdots \\
			{C{A^{f-2}}B}&{C{A^{f-3}}B}& \cdots &D
		\end{bmatrix}, \label{E5a}\\
		H_{f} &= \begin{bmatrix}
			{I}&{0}& \cdots &0\\
			{CK}&I& \cdots &0\\
			\vdots & \vdots & \ddots & \vdots \\
			{C{A^{f-2}}K}&{C{A^{f-3}}K}& \cdots &I
		\end{bmatrix}. \label{E5b}
	\end{align}
\end{subequations}
Past and future inputs are collected in the Hankel matrices
\begin{subequations} \label{E6}
	\begin{align}
		U_{p} &= \begin{bmatrix}
			{{u_{k-p}}}&{{u_{k-p+1}}}& \cdots &{{u_{k-p+N-1}}}\\
			{{u_{k-p+1}}}&{{u_{k-p+2}}}& \cdots &{{u_{k-p+N}}}\\
			\vdots & \vdots & \ddots & \vdots \\
			{{u_{k-1}}}&{{u_{k}}}& \cdots &{{u_{k+N-2}}} \end{bmatrix}, 	\label{E6b}\\
		U_{f} &= \begin{bmatrix}
			{{u_k}}&{{u_{k+1}}}& \cdots &{{u_{k+N-1}}}\\
			{{u_{k+1}}}&{{u_{k+2}}}& \cdots &{{u_{k+N}}}\\
			\vdots & \vdots & \ddots & \vdots \\
			{{u_{k+f-1}}}&{{u_{k+f}}}& \cdots &{{u_{k+f+N-2}}} \end{bmatrix}.  \label{E6a}\\
	\end{align}
\end{subequations}
The matrices $\Gamma_p$, $G_p$, $H_p$, $Y_p$, $Y_f$, $E_p$ and $E_f$ are defined in a similar way, see \cite{Qin2005novel}. The state sequences are defined as
\begin{subequations} \label{E7}
	\begin{align}
		X_{k} &= \begin{bmatrix}
			{{x_k}}&{{x_{k+1}}}& \cdots &{{x_{k+N-1}}} \end{bmatrix},\label{E7a}\\
		X_{k-p} &= \begin{bmatrix}
			{{x_{k-p}}}&{{x_{k-p+1}}}& \cdots &{{x_{k-p+N-1}}} \end{bmatrix}.\label{E7b}
	\end{align}
\end{subequations}
Although the state $X_k$ is unknown, it can be recovered from past inputs $U_p$ and outputs $Y_p$ using the relation
\begin{equation} \label{E8}
	X_k = L_pZ_p + \bar A^pX_{k-p},
\end{equation}
where $Z_p = \begin{bmatrix}
	Y_p^{\top}&U_p^{\top}
\end{bmatrix}^{\top}$ and $L_p$ is the extended controllability matrix of the predictor form \eqref{E2}, i.e., 
%\begin{equation} \label{E9}
%	L_p = \begin{bmatrix}
%		{\bar B}&{\bar A \bar B}& \cdots &{\bar A^{p-1}\bar B}
%	\end{bmatrix}.
%\end{equation}
\begin{equation} \label{E9}
	L_p = \begin{bmatrix}
		\bar A^{p-1}K& \cdots &{\bar A K}& K &{\bar A^{p-1}\bar B}&\cdots&{\bar A \bar B}&\bar B
	\end{bmatrix}.
\end{equation}
When $p$ is sufficiently large, $\bar A^p \approx 0$, hence, the bias term $\bar A^pX_{k-p}$ is omitted. Combining \eqref{E3a} and \eqref{E8}, we have
\begin{equation} \label{E10}
	Y_f = \Gamma_fL_pZ_p + G_fU_f + H_fE_f.
\end{equation}
Most SIMs use \eqref{E10} to first estimate the extended observability matrix $\Gamma_f$, and then obtain a realization of the system parameters up to a similarity transformation. To begin with, since $G_f$ is a lower-triangular Toeplitz matrix whose structure is difficult to be preserved with least-squares, this term is eliminated by projecting out $U_f$ as
\begin{equation} \label{E11}
	Y_f\Pi_{U_f}^{\perp} = \Gamma_fL_pZ_P\Pi_{U_f}^{\perp} + H_fE_f\Pi_{U_f}^{\perp},
\end{equation}
where $\Pi_{U_f}^{\perp} = I - U_f^\top(U_fU_f^\top)^{-1}U_f$. For the open-loop case, $U_f$ is uncorrelated with $E_f$, so we have $E_f\Pi_{U_f}^{\perp} \approx E_f$. Furthermore, since $E_f$ is uncorrelated with $Z_p$, i.e., $\frac{1}{N}E_fZ_p^\top \approx 0$, by multiplying $Z_p^\top$ on both sides of \eqref{E11} we have
\begin{equation} \label{E12}
	Y_f\Pi_{U_f}^{\perp}Z_p^\top \approx \Gamma_fL_pZ_P\Pi_{U_f}^{\perp}Z_p^\top.
\end{equation}
Now, the range space of the extended observability matrix $\Gamma_fL_p$ can be estimated by OLS, i.e.,
\begin{equation} \label{E13}
	\widehat {\Gamma_fL_p} = Y_f\Pi_{U_f}^{\perp}Z_p^\top(Z_P\Pi_{U_f}^{\perp}Z_p^\top)^\dagger,
\end{equation}
where $[\cdot]^\dagger$ is the Moore$\mbox{-}$Penrose pseudo-inverse. Then to recover the extended observability matrix $\Gamma_f$, weighted SVD is often used, i.e., 
\begin{equation} \label{E14}
	W_1\widehat {\Gamma_fL_p}W_2 = USV^\top \approx U_{n_x}S_{n_x}V_{n_x}^\top, 
\end{equation}
where $S_{n_x}$ contains the $n_x$ largest singular values. In this way, a balanced realization of $\Gamma_f$ is estimated as
\begin{equation} \label{E15}
	\hat \Gamma_f = U_{n_x}S_{n_x}^{\frac{1}{2}}.
\end{equation}
As pointed out in \cite{Van1995unifying}, one of the main differences among various classical SIMs regards the choice of weighting matrices used in the SVD-step. It is further pointed out in \cite{Gustafsson2002statistical} that $W_1$ has no influence on the asymptotic accuracy of the estimated observability matrix, and an approximately optimal weighting for $W_2$ is 
\begin{equation} \label{E16}
	W_2 = (Z_p\Pi_{U_f}^{\perp}Z_p^\top)^{1/2}. 
\end{equation}
There are two aspects to be mentioned:
\begin{enumerate}
	\item The lower-triangular Toeplitz matrix $G_f$ guarantees that the extended model is causal. Since it is eliminated during the projection step \eqref{E11}, the projected model becomes potentially non-causal. As a result, the estimated parameters have inflated variance due to the fact that unnecessary terms are included.
	\item For closed-loop data, $U_f$ and $E_f$ are correlated, i.e., $E_f\Pi_{U_f}^{\perp} \neq E_f$, so as a result, many methods are biased for closed-loop setups.
\end{enumerate}

To enforce causal models, a parallel PARSIM is proposed in \cite{Qin2005novel}. Instead of doing the projection once, PARSIM zooms into each row of \eqref{E10} and equivalently performs $f$ OLS projections. To illustrate this, the extended state-space model \eqref{E10} can be partitioned row-wise as 
\begin{equation} \label{E17}
	Y_{fi} = \Gamma_{fi}L_pZ_p + G_{fi}U_i + H_{fi}E_i, i = 1,2,...f,
\end{equation}
where 
\begin{equation} \label{E18}
	\begin{split}
		\Gamma_{fi} &= CA^{i-1},\\
		Y_{fi} &= \begin{bmatrix}
			{{y_{k+i-1}}}&{y_{k+i}}& \cdots &{y_{k+N+i-2}} \end{bmatrix},\\
		U_{fi} &= \begin{bmatrix}
			{{u_{k+i-1}}}&{u_{k+i}}& \cdots &{u_{k+N+i-2}} \end{bmatrix},\\
		E_{fi} &= \begin{bmatrix}
			{{e_{k+i-1}}}&{e_{k+i}}& \cdots &{e_{k+N+i-2}} \end{bmatrix},\\
		U_i &= \begin{bmatrix}
		{{U_{f1}^{\top}}}&{{U_{f2}^{\top}}}& \cdots &{{U_{fi}^{\top}}}
		\end{bmatrix}^{\top}, \\
		E_i &= \begin{bmatrix}
			{{E_{f1}^{\top}}}&{{E_{f2}^{\top}}}& \cdots &{{E_{fi}^{\top}}}
		\end{bmatrix}^{\top}, \\
		G_{fi} &= \begin{bmatrix}
			CA^{i-2}B&  \cdots &CB&D
		\end{bmatrix}  \overset{\Delta}{=}\begin{bmatrix}
			G_{i-1}&  \cdots &G_{1}&G_{0}
		\end{bmatrix}, \\
		H_{fi} &= \begin{bmatrix}
			CA^{i-2}K&  \cdots &CK&I
		\end{bmatrix} \overset{\Delta}{=} \begin{bmatrix}
			H_{i-1}&  \cdots &H_{1}&H_{0}
		\end{bmatrix}.
	\end{split}
	\nonumber
\end{equation}
Then the parallel PARSIM uses a bank of OLS to estimate $\Gamma_{fi}L_p$ and $G_{fi}$ simultaneously from the causal model \eqref{E17}:
\begin{equation} \label{E19}
	\begin{bmatrix}
		\widehat {\Gamma_{fi}L_p}& \hat G_{fi}
	\end{bmatrix} = Y_{fi} \begin{bmatrix}
		Z_p\\ U_i\end{bmatrix}^\dagger.
\end{equation}
At last, the whole estimate of $\Gamma_{f}L_p$ is obtained by stacking the $f$ values of $\widehat {\Gamma_{fi}L_p}$ together as
%\begin{equation} \label{E20}
%	\begin{bmatrix}
%		\widehat {\Gamma_{f1}L_p}\\ \widehat {\Gamma_{f2}L_p} \\ \vdots \\\widehat {\Gamma_{ff}L_p}\end{bmatrix} \overset{\Delta}{=} \widehat {\Gamma_{f}L_p}.
%\end{equation}
\begin{equation} \label{E20}
	\begin{bmatrix}
		\widehat {\Gamma_{f1}L_p} \\ \vdots \\\widehat {\Gamma_{ff}L_p}\end{bmatrix} \overset{\Delta}{=} \widehat {\Gamma_{f}L_p}.
\end{equation}
As we can see, by estimating the Markov parameters $G_{fi}$ in each row, the structure of the whole Toeplitz matrix $G_f$ is preserved. Furthermore, it is shown that regarding the estimates of the Markov parameters and the range space of the extended observability matrix, PARSIM generally gives smaller variance than conventional SIMs\citep{Qin2005novel}. However, to be consistent, PARSIM requires that there is no correlation between future $\{u_k\}$ and future $\{e_k\}$, which is only valid for the open-loop case. To make it applicable to the closed-loop case, an innovation estimation method is proposed in \cite{Qin2003closed}, which utilizes the structure of the Toeplitz matrix $H_f$ to decouple the correlation between future $\{u_k\}$ and $\{e_k\}$.

\subsection{SIMs Based on Predictor Form} \label{Sct2.3}
\vspace{-3mm}
To address the bias issue in classical SIMs when applied to closed-loop settings, several closed-loop SIMs have been put forward. In this section, we introduce two techniques that employ the predictor form \eqref{E2}, which provide inspiration for our method.

\subsubsection{SSARX} \label{Sct2.3.1}

By defining the following Toeplitz matrices for the predictor form \eqref{E2},
\begin{subequations} \label{E21}
	\begin{align}
		\bar \Gamma_f &= \begin{bmatrix}
			{{C^{\top}}}&{{{\left( {C\bar A} \right)}^{\top}}}& \cdots &{{{\left({C{\bar A^{f - 1}}} \right)}^{\top}}}
		\end{bmatrix}^{\top}, \\
		\bar G_{f} &= \begin{bmatrix}
			{D}&{0}& \cdots &0\\
			{C\bar B}&D& \cdots &0\\
			\vdots & \vdots & \ddots & \vdots \\
			{C{\bar A^{f-2}}\bar B}&{C{\bar A^{f-3}}\bar B}& \cdots &D
		\end{bmatrix}, \label{E5a}\\
		\bar H_{f} &= \begin{bmatrix}
			{0}&{0}& \cdots &0\\
			{CK}&0& \cdots &0\\
			\vdots & \vdots & \ddots & \vdots \\
			{C{\bar A^{f-2}}K}&{C{\bar A^{f-3}}K}& \cdots &0
		\end{bmatrix}, \label{E5b}
	\end{align}
\end{subequations}
we obtain the extended model
\begin{equation} \label{E22}
	\begin{split}
		Y_f &= \bar \Gamma_fX_k + \bar G_fU_f + \bar H_fY_f + E_f\\
		&= \bar \Gamma_fL_pZ_p + \bar G_fU_f + \bar H_fY_f + E_f.
	\end{split}
\end{equation}
To remove the possible correlation between $U_f$, $Y_f$ and $E_f$, SSARX \citep{Jansson2003subspace} first estimates the predictor Markov parameters $\left\{{C{\bar A^{i}}\bar B}\right\}$ and $\left\{{C{\bar A^i}K}\right\}$ from a high-order ARX model, and then replaces $\bar G_f$ and $\bar H_f$ with their estimates $\hat {\bar G}_f$ and $\hat {\bar H}_f$, leading to 
\begin{equation} \label{E23}
	Y_f - \hat {\bar G}_f U_f - \hat {\bar H}_f Y_f = \bar \Gamma_fL_pZ_p + E_f.
\end{equation}
Here, as $E_f$ is not correlated to $Z_p$, OLS can be used to obtain consistent estimates of $\bar \Gamma_fL_p$. Putting the truncation error of the high-order ARX model aside, as the estimates $\hat {\bar G}_f$ and $\hat {\bar H}_f$ are consistent for both open-loop and closed-loop data, SSARX is suitable for both cases.

\subsubsection{PBSID} \label{Sct2.3.2}
Inspired from SSARX, PBSID, also known as the whitening filter approach \citep{Chiuso2005consistency}, starts from the predictor form \eqref{E2} and utilizes the structure of the lower-triangular Toeplitz matrices $\bar G_{f}$ and $\bar H_{f}$ to carry out multi-stage projections row by row in \eqref{E22}. In this way, no pre-estimation as in SSARX is required, and causality is strictly enforced, similar to PARSIM. It should be further pointed out that PBSID is asymptotically equivalent to SSARX, in the sense of yielding the same asymptotic distribution of the estimators \citep{Chiuso2007role}. Later on, PBSID\textsubscript{opt}, an optimized PBSID algorithm, was proposed in \cite{Chiuso2007relation}. Besides sharing the advantages of PBSID, it also considers the distribution of the innovations $E_f$ and attains BLUE for estimating $\bar \Gamma_fL_p$.

\section{PARSIM\textsubscript{opt}} \label{Sct3}

In this section, we propose an optimized PARSIM algorithm. Just as PBSID\textsubscript{opt} refines PBSID, PARSIM\textsubscript{opt} preserves all the merits of PARSIM while demonstrating superior performance with reduced variance.
\vspace{-3mm}
\subsection{Weighted Least-Squares} \label{Sct3.1}
Our method stands on the shoulders of PARSIM. For the last term $H_{fi}E_i$ in \eqref{E17}, since $E_i$ is an unknown Hankel matrix and $H_{fi}$ contains unknown Markov parameters with respect to the innovations, the correlation structure of the noise in \eqref{E17} is not used in PARSIM, where OLS is used to estimate  $\Gamma_{fi}L_p$ and $G_{fi}$, which is consistent but generally not BLUE. Our contribution is to benefit from this structure to improve on the OLS estimate by way of WLS. For simplicity's sake, we now consider the SISO case, whereas the MIMO case can be dealt with similarly.

Taking $H_{fi} = \begin{bmatrix}
	H_1&H_0 \end{bmatrix}$ and $E_i = \begin{bmatrix}
	e_0&e_{1}\\
	e_{1}&e_{2}
\end{bmatrix}$, since $H_i$ and $e_k$ are scalars, we can rewrite $H_{fi}E_i$ as 
\begin{equation} \label{E24}
	H_{fi}E_i = \mathcal{E}_{i}\mathcal{T}_{\left[H_{fi}\right]},
\end{equation}
where $\mathcal{E}_{i}=\begin{bmatrix}
	e_0&e_{1}&e_{2} \end{bmatrix}$ is a row vector and $\mathcal{T}_{\left[H_{fi}\right]}$ is a Toeplitz matrix of $H_{fi}$, i.e.,
$\mathcal{T}_{\left[H_{fi}\right]} = \begin{bmatrix}
	H_1&0\\
	H_0&H_1\\
	0&H_0
\end{bmatrix}$. 
Substituting \eqref{E24} into \eqref{E17}, we have 
\begin{equation} \label{E25}
	Y_{fi} = \begin{bmatrix}
		\Gamma_{fi}L_p&G_{fi}\end{bmatrix} \begin{bmatrix}
		Z_p\\U_i\end{bmatrix}+\mathcal{E}_{i}\mathcal{T}_{\left[H_{fi}\right]}.
\end{equation}
In this way, there are no repeated elements in $\mathcal{E}_{i}$, and the problem can be fitted into the ML framework. To be specific, since $\mathcal{E}_{i} \sim \mathcal{N}(0,\sigma_e^2 I)$, we have $\mathcal{E}_{i}\mathcal{T}_{\left[H_{fi}\right]} \sim \mathcal{N}(0,\sigma_e^2 \mathcal{T}_{\left[H_{fi}\right]}^\top \mathcal{T}_{\left[H_{fi}\right]})$. With the optimal weighting matrix $W_i^{*} = (\mathcal{T}_{\left[H_{fi}\right]}^\top \mathcal{T}_{\left[H_{fi}\right]})^{-1}$, we have the BLUE in each step given by WLS as
\begin{equation} \label{E26}
	\begin{bmatrix}
		\widehat {\Gamma_{fi}L_p}& \hat G_{fi}
	\end{bmatrix} = Y_{fi}\begin{bmatrix}
	Z_p\\ U_i\end{bmatrix}^\top W_i^{*} \left(\begin{bmatrix}
	Z_p\\ U_i\end{bmatrix}W_i^{*}\begin{bmatrix}
	Z_p\\ U_i\end{bmatrix}^\top\right)^{-1}.
\end{equation}
Comparing with PARSIM that performs $f$ OLS to obtain $\widehat {\Gamma_{f}L_p}$, PARSIM\textsubscript{opt} performs $f$ WLS. Since the estimates $\widehat {\Gamma_{fi}L_p}$ of PARSIM\textsubscript{opt} have  smaller variances than PARSIM, the overall $\widehat{\Gamma_{f}L_p}$ of PARSIM\textsubscript{opt} has smaller variance. Furthermore, it has been shown that PARSIM generally gives smaller variance than other classical SIMs in terms of estimating the range space of the extended observability matrix. Thus, if the optimal weighting matrix is known, then PARSIM\textsubscript{opt} is BLUE and gives a smaller variance among those SIMs using the innovations form.

\begin{rem} \label{Rmk1}
	Notice that, in principle, \eqref{E24} can be omitted. However, to perform WLS directly on \eqref{E17} requires vectorizing the Hankel matrix $E_i$ resulting in a noise vector with repeated entries and thus a singular covariance matrix, which needs more careful investigations and computational efforts. This is the approach taken in PBSID\textsubscript{opt}.
\end{rem}

%\begin{rem} \label{Rmk4}
%	(Closed-loop Case) For closed-loop data, there is correlation between $U_f$ and $E_f$. If we directly use WLS or OLS, the result will be biased. As pointed out in SSARX, the estimates $\hat {\bar g}_i$ and $\hat {\bar h}_i$ from ARX model are not biased, even if for the closed-loop data. Thus, to fit in our framework, we can use a similar extraction algorithm to extract estimates of $\left\{C{A^{i}}B\right\}$ from $\left\{C{\bar A^{i}}\bar B\right\}$, then have an estimate of $\hat G_{fi}$. After that, we perform WLS for the following equation,
%	\begin{equation} \label{E31}
%		Y_{fi}-\hat G_{fi}U_i = \Gamma_{fi}L_pZ_p+\mathcal{E}_{i}\mathcal{T}\left[H_{fi}\right].
%	\end{equation}
%	As $\mathcal{E}_{i}$ is unrelated to $Z_p$, the result is consistent for closed-loop data. As we mentioned, there is an innovation estimation version of PARSIM, PARSIM\textsubscript{E}, which is consistent for the closed-loop case. It is interesting to compare PARSIM\textsubscript{E} with our method for closed-loop data, which we will show in the future.
%\end{rem}
\vspace{-3mm}
\subsection{Estimation of Weighting Matrices} \label{Sct3.2}

The preceding section assumes that the optimal weighting matrices $W_i^{*}$ are known, i.e., the terms $H_i = CA^{(i-1)}K$, $i = 1, 2,...,{f-1}$, are available,  which is generally not the case. To handle this, following the first step in SSARX, we propose to use a high-order ARX model to obtain consistent estimates of these Markov parameters. When the order $n$ is large enough, the ARX model captures the system dynamics with arbitrary accuracy. It is important to note that the order $n$ should be selected at a suitable rate with respect to the sample size $N$ in order to achieve consistency and asymptotic normality \citep{Ljung1992asymptotic}. Omitting the truncation error, consistent estimates of predictor Markov parameters can be obtained using OLS from the ARX model. Then, to form an estimate of the optimal weighting matrix $W_i^{*}$, we need to extract the estimates of $H_i = CA^{i-1}K$ from the estimates of $\bar H_i = C{\bar A^{i - 1}}K = C(A-KC)^{i - 1}K$, $i = 1,...,f$. Such extraction can be implemented by the following recursive algorithm \citep{Juang1993identification}:
\begin{subequations} \label{E27}
	\begin{align}
		H_1 & =  \bar H_1,\label{E27a}\\
		H_i & = \bar H_i + \sum_{j=1}^{i-1}{\bar H_j H_{i-j}}, {\text{for}} \: i \geq 2. \label{E27b}
	\end{align}
\end{subequations}
\vspace{-3mm}
\subsection{Estimation of System Parameters} \label{Sct3.3}
As we already mentioned in Section \ref{Sct2}, after obtaining $\widehat {\Gamma_{f}L_p}$, the next step for SIMs is to choose weighting matrices $W_1$ and $W_2$, and then perform SVD \eqref{E14}. For PARSIM\textsubscript{opt}, we choose the same weighting matrix as PARSIM, i.e., $W_1 = I$ and $W_2  = (Z_p\Pi_{U_f}^{\perp}Z_p^\top)^{1/2}$. As for the estimation of system matrices, it follows similar steps as PARSIM. To conserve space, these steps are omitted, and we refer readers to \cite{Qin2005novel} for more details. 
\vspace{-3mm}
\section{Simulation Examples} \label{Sct4}
Here we provide some numerical examples to compare PARSIM\textsubscript{opt} with state-of-the-art. For implementations of the methods included in the comparison, we use the following resources: N4SID with CVA weighting \citep{Van2012subspace}, SSARX \citep{Jansson2003subspace}, PBSID\textsubscript{opt} \citep{Chiuso2007relation} and PARSIM \citep{Qin2005novel}. Here are some common settings of the first two examples:
\begin{enumerate}
	\item The matrix $D$ is constrained to be 0 in all methods.
	\item The comparison is under the open-loop case.
	\item The future horizon $f$ is fixed, and the past horizon $p$ is selected according to the AIC criterion \citep{Akaike1974new} for all methods.
	\item PEM as implemented in the Matlab (R2021a) System Identification Toolbox is used as a benchmark.
\end{enumerate}
 \vspace{-3mm}
\subsection{Example 1} \label{Sct4.1}

\begin{figure}
	\centering
	\includegraphics[scale=0.6]{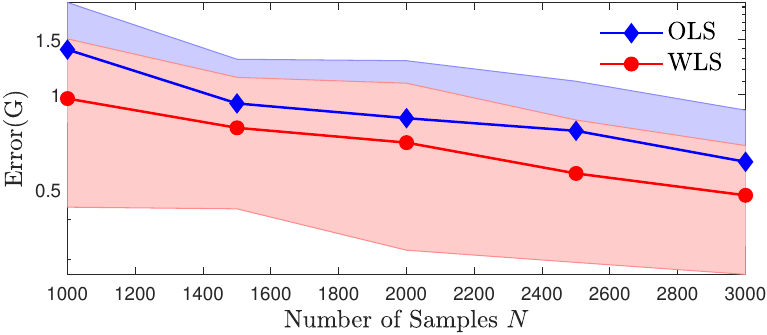}
	\caption{Errors of Markov parameters.}
	\label{F1}
\end{figure}

The first example is used in \cite{Jansson2003subspace}, which is
\begin{equation*}
	y_t = \frac{0.21q^{-1}+0.07q^{-2}}{1 - 0.6q^{-1} + 0.8q^{-2}} u_t + \frac{1}{1 - 0.98 q^{-1}} e_t,
\end{equation*}
where $u_{k} \sim \mathcal{N}(0,1)$ and $e_{k} \sim \mathcal{N}(0,4)$. 
We first show that WLS in PARSIM\textsubscript{opt} gives smaller variance than OLS in PARSIM when estimating the Markov parameters and the range space of the extended observability matrix. For this purpose, we consider sample sizes $N = 1000:500:3000$ and $f=10$, and the performance is evaluated by 
\begin{equation*}
	{\rm{Error}}(G) = \|\hat G_{ff} - G_{ff}\|/\|G_{ff}\|,
\end{equation*}
where $G_{ff}$ is defined in \eqref{E17}, and $\|G_{ff}\|$ is the 2-norm of $G_{ff}$. We remark that here $\hat G_{ff}$ is directly obtained by solving WLS or OLS in \eqref{E25}, not from the identified system matrices $A,B$ and $C$ indirectly. Figure \ref{F1} shows the estimation error of the Markov parameters, where the solid lines indicate average errors of each method in 50 Monte Carlo trials, and the shadowed bars around them mean the variances in 50 trials. It is clear that WLS in PARSIM\textsubscript{opt} outperforms OLS in PARSIM.

\begin{figure}
	\centering
	\includegraphics[scale=0.6]{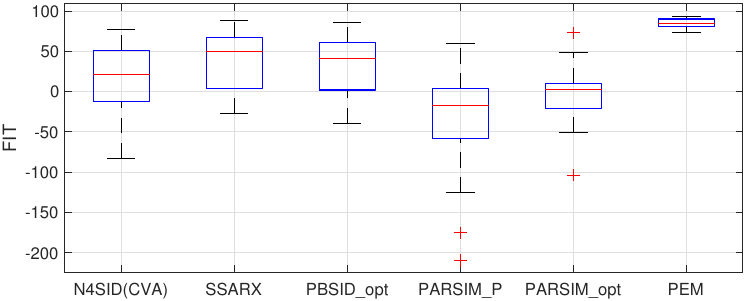}
	\caption{FIT (Example 1).}
	\label{F2}
\end{figure}

Second, we show the comparison between PARSIM\textsubscript{opt} and other SIMs. The performance is evaluated by 
\begin{equation*}
	{\rm{FIT}} = 100\left(1-\frac{\|g_{o} - \hat g\|}{\|g_{o} - {\rm{mean}}[g]\|}\right),
\end{equation*}
where $g_{o}$ is a vector with the impulse response parameters of the true transfer function from $u$ to $y$, and similarly for $\hat g$ but for the estimated model of different SIMs. The sample size $N = 2000$, and the FIT result based on 50 Monte Carlo simulations is shown in Figure \ref{F2}. We remark that, for a clear representation, outliers are removed from the boxchart. As we can see, for this example SIMs based on innovations form (N4SID and PARSIM) are not as good as SIMs based on the predictor form (SSARX and PBSID\textsubscript{opt}). In addition, PARSIM\textsubscript{opt} is better than PARSIM.
\vspace{-3mm}
\subsection{Example 2} \label{Sct4.2}
\begin{figure}
	\centering
	\includegraphics[scale=0.6]{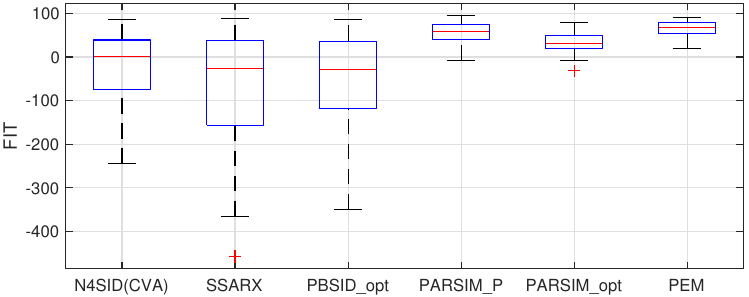}
	\caption{FIT (Example 2).}
	\label{F3}
\end{figure}

The second example is used in \cite{Jansson1998consistency} and \cite{Qin2005novel}. For this counterexample, many SIMs are not consistent and have a poor performance. The system is given by
\begin{equation*}
	\begin{split}
		x_{k+1} &= \begin{bmatrix}
			2\gamma&-\gamma^2 \\ 1&0\end{bmatrix}x_{k} +  \begin{bmatrix}
			1 \\ 
			-2 \end{bmatrix}u_{k} + \begin{bmatrix}
			-0.21 \\ 
			-0.559 \end{bmatrix}e_{k}, \\
		y_{k} &= \begin{bmatrix}2&-1\end{bmatrix}x_{k} + e_{k},
	\end{split}
\end{equation*} 
where $\gamma = 0.9184$, $e_{k} \sim \mathcal{N}(0,217.1)$, 
$u_k = (1-\gamma q^{-1})^2(1+\gamma q^{-1})^2r_{k}$ and $r_{k} \sim \mathcal{N}(0,1)$. The sample size $N = 2000$ and $f=7$, and the FIT result based on 50 Monte Carlo simulations is shown in Figure \ref{F3}. As we can see, for this example SIMs based on innovations form (N4SID and PARSIM) are better than SIMs based on the predictor form (SSARX and PBSID\textsubscript{opt}). In addition, PARSIM\textsubscript{opt} is slightly worse than PARSIM. The reason is that due to the large variance in innovations, the pre-estimation step is not good, which results in a bad estimate of the optimal weighting matrix for PARSIM\textsubscript{opt}.
%\subsection{Example 3: Random systems} \label{Sct4.3}
%
%In order to test the robustness of PARSIM\textsubscript{opt}, we now perform a simulation with fourth order random systems generated by MATLAB as follows:
%\begin{equation*}
%	\begin{split}
%		&{\rm{m0 = idss(drss(4,1,1));}}\\
%		&{\rm{m0.d = zeros(1,1);}}\\
%		&{\rm{u = randn(2000,1);}}\\
%		&{\rm{y = sim(m0,u) + sqrt(10*rand())*randn(2000,1);}}	
%	\end{split}
%\end{equation*}
%Besides, according to a critical view on the MATLAB command 	${\rm{drss}}$ \citep{Rojas2015critical}, the magnitude of the sampled dominant pole $p_{\rm{max}}$ is restricted to satisfy $0.78<p_{\rm{max}}<0.9$.  As suggested in \cite{Qin2006overview}, the optimal past and future horizons for SIMs based on innovations form and predictor form are different. With this consideration, we create a candidate set for $f = 6:2:20$, and $p$ is selected according to the AIC criterion. At last, we choose the optimal $f$ and $p$ that minimize the prediction error for each method. Then the optimal FIT result of each method based on 50 Monte Carlo simulations is shown in Figure \ref{F4}. As we can see, PARSIM\textsubscript{opt} gives the best performance among those methods. Thus, by combining features of both the innovations form and the predictor form, PARSIM\textsubscript{opt} is a fairly safe choice for users.
%
%\begin{figure}
%	\centering
%	\includegraphics[scale=0.6]{drss_siso_fit}
%	\caption{FIT (Fourth order random systems).}
%	\label{F4}
%\end{figure}
\vspace{-3mm}

\subsection{Example 3: Random systems} \label{Sct5.3}
\begin{figure}
	\centering
	\includegraphics[scale=0.55]{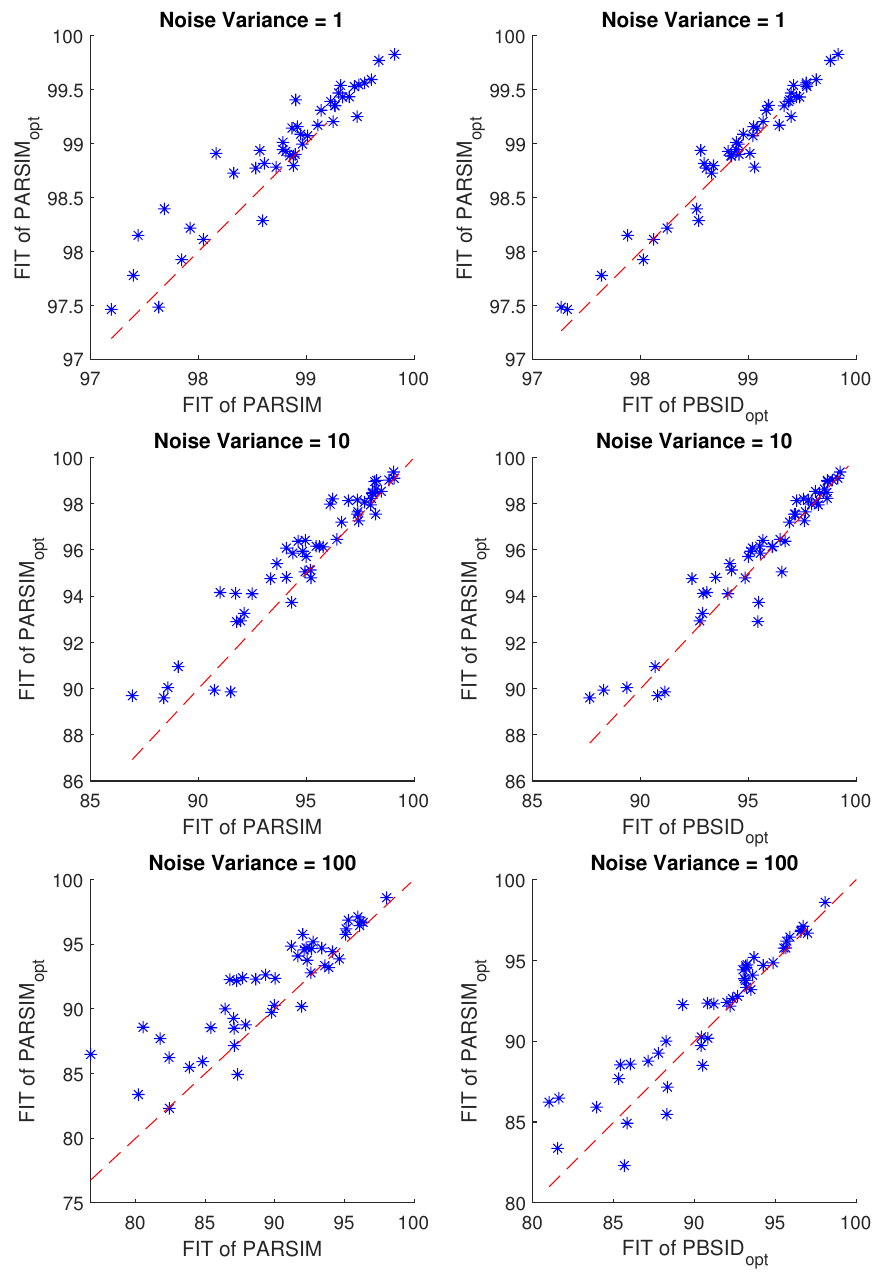}
	\caption{Joint FIT distribution of random systems.}
	\label{F5}
\end{figure}

In order to test the robustness of PARSIM\textsubscript{opt}, we now perform a simulation with sixth order random systems generated by MATLAB as follows:
\begin{equation*}
	\begin{split}
		&{\rm{m = idss(drss(6,1,1));}}\\
		&{\rm{m.d = zeros(1,1);}}{\rm{m.b = 5*randn(6,1);}}\\
		&{\rm{u =idinput([1000,1],'rbs',[0\;0.1]);}}\\
		&{\rm{y = sim(m,u) + \sigma_e*randn(1000,1);}}	
	\end{split}
\end{equation*}
where $\sigma_e^2$ is the noise variance. In addition, according to a critical view on the MATLAB command ${\rm{drss()}}$ \citep{Rojas2015critical}, the magnitude of the sampled dominant pole $p_{\rm{max}}$ is restricted to satisfy $0.78<p_{\rm{max}}<0.9$. As it has been shown that PBSID is asymptotically equivalent to SSARX, and PARSIM generally gives smaller variance than classical SIMs \cite{Chiuso2007role}, we mainly evaluate the performance of PARSIM\textsubscript{opt} with respect to PARSIM and PBSID\textsubscript{opt}. We believe that such comparison is interesting to the field, as PARSIM and PBSID algorithms both partition the extended state-space model row-wise and estimate a bank of ARX models using multi-step least-squares. Also, in terms of estimating the range space of the extended observability matrix, PBSID\textsubscript{opt} is BLUE among those methods based on the predictor form, while now  PARSIM\textsubscript{opt} is BLUE among those methods based on the innovations form. For a fair comparison, we use three different noise levels, i.e., $\sigma_e^2 = 1,10,100$. For each noise level, 50 independent random systems are generated. The joint FIT distribution of PARSIM\textsubscript{opt} with respect to PARSIM and PBSID\textsubscript{opt} are presented in Figure \ref{F5}, where each blue $*$ marker means a random system, and the red line is a bisector line. As shown in the left column of Figure \ref{F5}, the improvement of PARSIM\textsubscript{opt} over PARSIM is quite clear, as for different noise levels, PARSIM\textsubscript{opt} has higher FIT than PARSIM on most systems. As shown in the right column, when the noise level is small, the performance of PARSIM\textsubscript{opt} and PBSID\textsubscript{opt} is roughly equal, and when the level of noise increases, PARSIM\textsubscript{opt} appears to perform more effectively than PBSID\textsubscript{opt}.
\vspace{-3mm}

\section{Conclusion} \label{Sct5}
\vspace{-3mm}
Standing on the shoulders of PARSIM and SSARX, this paper presents 
an alternative way to handle processing preceding the SVD-step in subspace identification. The idea is to obtain a more accurate estimate of the observability matrix by replacing ordinary least-squares with weighted-least squares, where the optimal weighting matrix is substituted by a consistent estimate thereof. While the method follows PARSIM, the estimate of the optimal weighting matrix is based on a high-order ARX model, thus borrowing features from SSARX. Simulation examples suggest that this mixed use of both the innovations form and the predictor form in the estimation method is beneficial for the final state-space model estimate. In the future, we will implement PARSIM\textsubscript{opt} for MIMO systems, and extend it to the closed-loop case. Meanwhile, more simulations will be done to reveal its potential benefits.

\begin{ack}
The authors would like to thank Alessandro Chiuso and Magnus Jansson for insightful discussions and sharing codes of PBSID and SSARX, respectively.
\end{ack}

\bibliography{ifacconf}             % bib file to produce the bibliography

%\appendix
%\section{A summary of Latin grammar}    % Each appendix must have a short title.
%\section{Some Latin vocabulary}              % Sections and subsections are supported  
                                                                         % in the appendices.
\end{document}